\documentclass[fleqn,usenatbib]{mnras}

\usepackage{newtxtext,newtxmath}

\usepackage[T1]{fontenc}
\usepackage{ae,aecompl}

\usepackage{graphicx}	
\usepackage{amsmath}	

\newcommand{\msun}{M$_{\odot}$}
\newcommand{\sigmagas}{$\Sigma_\text{gas}$}
\newcommand{\sigmadust}{$\Sigma_\text{dust}$}
\newcommand{\deltadust}{$\delta_\text{dust}$}
\newcommand{\deltagas}{$\delta_\text{gas}$}
\newcommand{\gasdust}{$\Delta_\text{g/d}$}
\newcommand{\rsub}{$R_\text{sub}$}
\newcommand{\rgap}{$R_\text{gap}$}
\newcommand{\rcav}{$R_\text{cav}$}
\newcommand{\stellarPH}{$\log$(P/H)$_{\star}$}
\newcommand{\kms}{km\,s$^{-1}$}
\newcommand{\loggf}{$\log{(g_{l}f_{lu})}$}

\newcommand{\resub}[1]{#1}

\title[Phosphorus in HD\,100546]{Refractory phosphorus in the HD\,100546 protoplanetary disk}

\author[M. Kama et al.]{
Mihkel Kama$^{1,2}$\thanks{E-mail: m.kama@ucl.ac.uk},
Oliver Shorttle$^{3}$,
Sandipan P. D. Borthakur$^{2}$,
Luke Keyte$^{1, 4}$,
\newauthor Jennifer B. Bergner$^{5}$, Luca Fossati$^{6}$, Colin P. Folsom$^{2}$, Heleri Ramler$^{2}$
\\
$^{1}$Department of Physics and Astronomy, University College London, Gower Street, London, WC1E 6BT, UK\\
$^{2}$Tartu Observatory, University of Tartu, Observatooriumi 1, T\~{o}ravere, 61602, Estonia\\
$^{3}$Institute of Astronomy, University of Cambridge, Madingley Road, Cambridge, CB3 0HA, UK\\
$^{4}$Astronomy Unit, School of Physics and Astronomy, Queen Mary University of London, London E1 4NS, UK\\
$^{5}$University of California, Berkeley, Department of Chemistry, Berkeley, CA 94720, USA\\
$^{6}$Space Research Institute, Austrian Academy of Sciences, Schmiedlstrasse 6, 8042 Graz, Austria\\
}

\date{Accepted XXX. Received YYY; in original form ZZZ}

\pubyear{2021}

\begin{document}
\label{firstpage}
\pagerange{\pageref{firstpage}--\pageref{lastpage}}
\maketitle

\begin{abstract}
The phosphorus budget of planets is intertwined with their formation history and is thought to influence their habitability. The chemical reservoirs and volatile \emph{vs} refractory budget of phosphorus in planet-forming environments have so far eluded empirical characterisation. We employ high-resolution spectra from HST/STIS in the ultraviolet and APEX in the sub-mm to constrain the phosphorus budget in the well-characterized HD\,100546 star and protoplanetary disk system. We measure $\log{(P/H)_{\star}}=-7.50^{+0.23}_{-0.28}$ on the stellar surface, which traces the total inventory of P in accreting gas \emph{and }dust from the inner disk. The inner disk gas, inside of the main dust trap, has $\log{(P/H)_{\rm in}}\lesssim-8.70$, and the outer disk gas $\log{(P/H)_{\rm out}}\lesssim-9.30$. \resub{Phosphorus in the disk is carried by a relatively }refractory reservoir, consistent with minerals such as apatite or schreibersite\resub{, or with ammonium phosphate salts,} in terms of sublimation temperature. \resub{We discuss the impact this might have on the two protoplanets around HD\,100546.} Our results contribute to our understanding of the chemical habitability of planetary systems and lay a foundation for future explorations, especially in the context of JWST and \emph{Ariel} which can study phosphorus in exoplanet atmospheres.
\end{abstract}




\section{Introduction}

How phosphorus (P) arrives in planets, and in what abundance, is of particular importance across astrophysics, Earth, and planetary sciences.  The phosphorus abundance in the envelopes of giant planets constitutes one of the clues to their formation history \citep{ObergWordsworth2019}. P also has an essential role in biology, in the information-carrying molecules of life \citep{westheimer1987_nature}, and prebiotic chemistry \citep[as a buffer and reagent;][]{patel2015_natchem}.  As the debate around the phosphine detection in Venus's atmosphere has illustrated \citep[e.g.,][]{greaves2021_nast,villanueva2021_nast,lincowski2021_apjl}, gas-phase phosphorus also has potential as a biosignature, albeit this interpretation is critically dependent on the magnitude and chemistry of abiotic exogeneous and endogeneous sources \citep{sousa2020_astrobiology,bains2021_astrobiology}.  The phosphorus chemistry of planetary atmospheres and surfaces is therefore a key consideration in the search for life on exoplanets in the era of JWST, Ariel, and high-resolution ground-based spectroscopy.

Phosphorus exhibits a diverse behaviour in astrophysical and exoplanetary contexts: \resub{it }undergoes a journey from its formation in stars, to the diffuse and dense interstellar medium, to planets that may take it from gas, to solid, and potentially back to gas again, whilst transitioning across a wide range of oxidation states \citep[from -3 to +5; e.g., ][]{jenkins2009_apj,pasek2007_2007}. This rich chemical behaviour means there is still much to be understood about the distribution and chemistry of phosphorus in planetary systems.  Developing this understanding is ultimately a foundation to prospecting the universe for life --- phosphorous is a key part of chemical habitability \citep{Krijtetal2023}.

Phosphorus begins its journey to planets in the interstellar medium, having been synthesised and delivered by dying massive stars. It is usually classed as a (semi-)refractory element, locked in solids at equilibrium temperatures $\leq1229\,$K \citep{Lodders2003}, but observations of the diffuse atomic interstellar medium indicate that all elemental P is accounted for by \emph{atomic gas} in the most diffuse ISM regions \citep[][and references therein]{Jenkins2009}. The gas-phase phosphorus budget in the molecular interstellar medium is thought to be dominated by three species: PO, PN, and HCP \citep{Mininnietal2018,Chantzosetal2020}, with the latter being mainly seen around evolved stars. PO and PN are easily synthesized in the gas phase \citep{Thorneetal1984, Millaretal1987}. Progressive improvements to the phosphorus chemical network yield the same overall conclusion regarding the main gas-phase reservoirs \citep{Adamsetal1990, Millar1991, MacKayCharnley2001, AotaAikawa2012, JimenezSerraetal2018, Chantzosetal2020,Siletal2021,FernandezRuzetal2023}.

Since the first detection of PN in Orion\,KL and other high-mass star-forming regions \citep{Ziurys1987}, PN, PO, and HCP have been found in various combinations in cold and warm star-forming cores \citep{Fontanietal2016, Rivillaetal2016, Mininnietal2018, Wurmser2022}, Galactic Centre clouds \citep{Rivillaetal2018}, and in outflows from protostars \citep{Yamaguchietal2011, Leflochetal2016, Bergner2019} and evolved stars \citep{Milametal2008, DeBecketal2013}. Additionally, HCP and PH$_{3}$ have been found in the outflowing envelope around the evolved star IRC$+$10216 \citep{Agundezetal2007, Agundezetal2008, Agundezetal2014b}.

The above observational studies find gas-phase PO, PN, and HCP abundances spanning three orders of magnitude: from as low as $\rm X/H_{2}=10^{-10}$ to as high as the solar \resub{elemental P/H }value $2.5\times 10^{-7}$ \citep{Asplundetal2009}. These values correspond to a range of depletions from strong, factor of $1000$ depleted gas-phase elemental phosphorus, to undepleted within the uncertainties.  The depletion of P from the gas phase is highly environment-dependent, with nearly all elemental phosphorus in the gas in the diffuse ISM \citep{Jenkins2009}. Observations of star-forming regions lend support to the importance of sputtering in keeping P-species in the gas phase\resub{, because diatomic P-bearing molecules are abundant in outflow-shocked material where solids are collisionally eroded} \citep{Mininnietal2018, Wurmser2022}. In terms of thermal desorption, PO and PN have a desorption energy of $E_{\rm B}=1900\,$K or $5770\,$K \citep[$T_{\rm sub}\approx 35\,$K or $125\,$K;][]{Rivillaetal2016, piacentino_2022}, placing PO and PN in the domain of (hyper-)volatiles. Phosphine (PH$_{3}$) may be an important P-bearing ice species, but models suggest that in photon-dominated regions, such as the observable outer layers of protoplanetary disks, any PH$_{3}$ released into the gas is photodissociated; the released P then goes on to form PO and PN \citep{Rivillaetal2020}. \resub{There is a lack of gas-phase PN or PO towards hot core environments \citep[e.g.][]{Bernal2021, Fontani2024}. Hot cores are the centrally irradiated, ``hot'' ($T_{\rm kin}\gtrsim 100\,$K) inner regions of protostellar cores \citep{Wilsonetal1979} where volatile ices sublimate \citep{GarrodWidicusWeaver2013} and should be detectable in the gas. It is therefore unlikely that grain-phase P is mainly in a volatile ice such as PN, PO, or PH$_3$.} Instead, these molecules are likely formed in the gas after shock sputtering of a semi-volatile grain carrier like apatite minerals \citep{Bergner2022}.

Thermodynamic models have shown that P condensation in a disk is highly sensitive to pressure, assumed gas-phase species, and the details of condensation \citep{Sears1978, FegleyLewis1980}. Equilibrium condensation models for early solar system conditions indicate that phosphorus first transfers from the gas into the mineral schreibersite \citep[with a half mass condensation temperature $T_{\rm 50\%}\sim1229\,$K at solar composition;][]{Lodders2003}, before transforming to phosphate minerals at lower temperatures \citep[$T<800$\,K;][]{pasek2019_icarus}.  

Whilst in the solar system we have a record of the P-bearing condensation products \resub{and their altered forms} in the form of primitive CI chondrites containing schreibersite and apatite  \citep[e.g.,][]{Morloketal2006}, \resub{the poorly known initial conditions and chemical evolution of disks and planetesimals mean }these only give us a partial insight into how phosphorus \resub{generally behaves in planet-forming environments}.  Our new observations of phosphorus in \resub{a planet-forming disk and its accretion-dominated host star} will fill this gap, and directly probe its gas-solid phase partitioning. 

Stars earlier than type F5 (M$_{\star}\gtrsim1.4\,$M$_{\odot}$) do not have deep convetive mixing in their envelope, rather being dominated by rotational mixing and diffusion \citep{Turcotte2002}. This allows accretion from a protoplanetary disk to ``contaminate'' the observable photosphere \citep{JermynKama2018}. The abundance of dust-forming elements has been shown to be low on the surface of Herbig\,Ae/Be stars that have strong dust traps (visible as bright dust rings) in their disks, the so-called ``transitional disks'' \citep{Kamaetal2015b,GuzmanDiazetal2023}. The presence of a strong dust trap removes refractory elements from the accretion stream, and this provides a diagnostic method to measure the refractory fraction of volatile and semi-volatile elements. We have previously used this method to determine the refractory fraction of sulfur \citep[$89\pm8\,$\% locked in refractory solids;][]{Kamaetal2019}.

The depletion behaviour of phosphorus from the gas phase is not yet fully understood, so observations of P-carriers in physically well-characterised environments are needed to more accurately model its partitioning between specific gas- and solid-phase species before, and during, planet formation. In this paper, we apply the stellar accretion contamination method for the first empirical study of the major phosphorus carriers in a protoplanetary disk. This specifies the target selection criteria: an intermediate-mass young star with a transitional protoplanetary disk, both with well-characterised bulk parameters and as much information on chemical abundances as possible.


\section{The HD 100546 system}
\label{sec:target}


The Herbig\,Ae/Be system HD\,100546 consists of a well-characterised A0-type star ($2.4\,$M$_{\odot}$) surrounded by a warm, flaring transitional disk. \resub{The large grains in the disk are trapped in an inner ($22$ to $40\,$au) and outer ($150$ to $230\,$au) dust ring. }The \resub{outer dust gap and inner cavity each host a protoplanet candidate: }HD\,100546\,b \citep[$a_{\rm maj,b}=143\,$au, $M_{\rm b}=3\,$M$_{\rm Jup}$][]{} and c \citep[$a_{\rm maj,b}=13\,$au,$M_{\rm b}=8\, $M$_{\rm Jup}$][]{Pinillaetal2015, Pyerinetal2021}. There is a significant literature on the source, we direct the reader to \citet{Benistyetal2010}, \citet{Panicetal2010}, \citet{Brudereretal2012}, \citet{Quanzetal2013}, \citet{Walshetal2014}, \citet{Pinillaetal2015}, \citet{Kamaetal2016b}, \citet{BoothAetal2021_HD100}, \citet{BoothAetal2024}, and \citet{Keyteetal2023, Keyteetal2024} as a starting point. The physical and chemical structure of the disk has been modelled in detail, reproducing observed rotational line emission from the main carbon-, oxygen-, and sulfur-carrying species. The temperature structure of the disk is such that the direct freeze-out of gas-phase hypervolatiles (e.g., CO) in the disk is negligible \citep{Brudereretal2012, Kamaetal2016b, PanicMin2017}.  The total volatile elemental C and O are not substantially depleted from the gas phase \citep{Brudereretal2012, Kamaetal2016b}, while sulfur is depleted by a factor $\sim1000$ with large radial variation \citep{Keyteetal2023, Keyteetal2024}. The gas-phase abundances of C, O, and S vary radially and azimuthally \citep{Keyteetal2023, Keyteetal2024}, implying the chemical composition of material available to accrete may differ for the two protoplanets.



\section{Observations}

To determine the budget of refractory and volatile phosphorus carriers in the planet-forming disk around HD\,100546 (see Section\,\ref{sec:target}), we study emission in rotational transitions of P-bearing molecules (PO, PN, and HCP) in the disk gas and atomic absorption lines of P ionization states in the photosphere of the central star.

\subsection{HST/STIS}\label{sec:hstdata}

To determine the stellar photospheric phosphorus abundance, we used archival spectra from the Space Telescope Imaging Spectrograph (STIS) on \emph{Hubble} (HST). The high-resolution (R\,$\approx114\,000$) data \resub{covered wavelengths from $1150$ to $2887\,$\AA\ and }was downloaded from the StarCAT catalogue (see Data Availability). Two spectra of the star were available, taken on 2000-07-22 and 2000-11-04, with signal-to-noise SNR${\approx}10$ and ${\approx}30$, respectively, at ${\approx}1500\,\Angstrom$. We used the spectrum with the higher SNR.

\resub{The use of the HST/STIS spectrum to derive the stellar P/H ratio for HD\,100546 is described in Section\,\ref{sec:stellarPH}.}

\begin{table}
\caption{APEX sub-mm spectroscopic observations of HD\,100546.}
\centering
\begin{tabular}{c c c c}
\hline\hline
Date & $\nu_{\rm LO}$ & $t_{\rm exp}$ & PWV\\
{yyyy-mm-dd} & [GHz] & [hours] & [mm] range \\
\hline
2018-07-30 & $234$ & $2.4$ & $0.9\ldots1.2$\\
    & $234$, $242$ & $3.6$ & $1.2\ldots1.6$ \\
2018-07-31 & $234$, $242$ & $4.8$ & $1.0\ldots2.2$ \\
2018-08-03 & $242$ & $2.8$ & $1.6\ldots1.9$ \\
2018-11-06 & $242$ & $2.5$ & $1.2\ldots1.4$ \\
2018-08-08 & $242$ & $3.9$ & $2.6\ldots2.9$ \\
2018-11-07 & $234$, $242$ & $2.8$ & $0.8\ldots1.6$\\
2018-11-12 & $234$ & $4.2$ & $3.2\ldots3.9$\\
\hline
\end{tabular}
\\
\flushleft
\emph{Notes. }The columns give the date (1), local oscillator frequency setting $\nu_{\rm LO}$ (2), on-source exposure time (3), and precipitable water column (4). Grouped LO settings were observed in the same observing window on a given night.
\label{tab:apexobs}
\end{table}

\subsection{APEX}\label{sec:apexdata}

To study molecular emission lines of phosphorus-bearing species in the protoplanetary disk, observations of PO and PN rotational lines were carried out with the PI230 receiver and FFTS-4G backend on the APEX sub-millimetre telescope in 2018 (proposal E-0102.C-0927A). Two local oscillator settings were used: $\nu_{\rm LO}=234\,$GHz for a set of PO and HCP lines and $\nu_{\rm LO}=242\,$GHz for PN. A summary of the sub-millimetre observations is given in Table\,\ref{tab:apexobs}.

\resub{The APEX data underwent standard reduction procedures at ESO. After applying minor, low-order baseline corrections, we did not detect any of the targeted lines. The observing settings were chosen in such a way as to ensure that even upper limits would provide useful constraints on the gas-phase phosphorus budget, which we discuss in Section\,\ref{sec:diskPH}.}

\begin{table}
\caption{PO, PN, and HCP rotational transitions observed with APEX.}
\centering
\begin{tabular}{c c c c c}
\hline\hline
Mol. & Transition & $\nu_{\rm ul}$ & RMS & $T_{\rm A}\,dv$\\
 & & (GHz) & (mK) & (mK\,$\frac{\rm km}{\rm s}$)\\
\hline
PN & $J=5$--$4$ & $234.936$ & $2.8$ &  <13.2\\
HCP & $J=6$--$5$ & $239.694$ & $2.8$ &  <26.4\\
PO & $J=\frac{11}{2}$--$\frac{9}{2}, F=6$--$5,l=e$ & $239.949$ & $2.8$ & <26.4\\
PO & $J=\frac{11}{2}$--$\frac{9}{2}, F=5$--$4, l=e$ & $239.958$ & $2.8$ & <26.4\\
PO & $J=\frac{11}{2}$--$\frac{9}{2}, F=6$--$5, l=f$ & $240.141$ & $2.5$ & <23.7\\
PO & $J=\frac{11}{2}$--$\frac{9}{2}, F=5$--$4, l=f$ & $240.153$ & $2.6$ & <24.6\\
\hline
\end{tabular}
\\
\flushleft
\emph{Notes. }Upper limits are given with $3\sigma$ confidence. The HCP line was not included in the modelling.
\label{tab:apexlines}
\end{table}

\section{Analysis}

We investigate the phosphorus budget in the HD\,100546 system by measuring its abundance relative to hydrogen on the stellar surface and in the disk gas. With the stellar surface measurement, we constrain the total amount of phosphorus passing through the last major dust trap and accreting onto the star; while, with the disk gas measurement, we constrain the fraction of elemental P locked in either dust grains or midplane ices.

As a reference value, we adopt the solar abundance of phosphorus, $\log{\rm(P/H)}_{\odot}=-6.59\pm0.03$ \citep{Asplundetal2009}. An alternative would be the proto-solar value, but estimates differ significantly: \citet{Lodders2003} estimated $\log{\rm(P/H)_{\odot}^{\rm prot}}=-6.46\pm0.04$, while \citet{Asplundetal2021} suggest $\log{\rm(P/H)_{\odot}^{\rm prot}}=-6.526$. We use the current solar abundance to place constraints on the unseen solid component. The abundances are conventionally expressed as the base-10 logarithm of the elemental number ratio \stellarPH\ for stellar photospheres, and as the absolute ratio for disk gas.

The disk around HD\,100546 includes a bright inner dust ring from $22$ to $40\,$au, and a faint outer ring from $150$ to $230\,$au. We can assume a fraction $\delta_{\rm L}$ of the inward-moving dust mass is trapped as large grains in the inner ring dust trap. This removes a corresponding fraction of the refractory component of each chemical element from each parcel of disk material. Thus, if $100\,$\% of Fe in a parcel of disk material is contained in refractory dust, then $100\times(1-\delta_{\rm L})\,$\% of all Fe would pass through the main dust trap and accrete onto the star, where it can be observed in the stellar photospheric composition as long as the mixing is slow and the accretion rate is high enough \citep{Turcotte2002, JermynKama2018}. This method allows to infer the fraction of each chemical element locked in solid particles. We apply it below to phosphorus.

The high disk accretion rate onto HD\,100546, $\log(\dot{M}_{\rm acc})=-7.04_{-0.15}^{+0.13}\,$M$_{\odot}\,$yr$^{-1}$ \citep{Fairlambetal2015}, leads the photospheric mixing fraction of freshly accreted material to be $f_{\rm acc}=98\,$\%\resub{, using a stellar mixing model that includes rotation, diffusion, and other processes} \citep{JermynKama2022}. The stellar photospheric composition is thus, to a high degree of accuracy, the elemental composition of the disk material \emph{currently} accreting onto the star.

\subsection{The accreted stellar phosphorus abundance}\label{sec:stellarPH}

\resub{To measure the stellar photospheric P/H ratio, we }adopted stellar parameters and known photospheric abundances for HD\,100546 from \citet{Kamaetal2016b}, who followed the spectral synthesis and fitting methodology of \citet{Folsometal2012}. \resub{The 2016 }analysis did not include phosphorus due to a lack of suitable lines in \resub{the ESO FEROS spectrograph }data. This is not unusual: because of a lack of suitable lines, the phosphorus abundance is rarely \resub{measured in intermediate-mass stars. Lines of P are however present in the HST/STIS spectrum we used. In our analysis of the STIS data, we employ }the Zeeman spectral synthesis code \citep{Landstreet..1988,Wade..2001,Landstreet..SiriusA..2011, Folsometal2012}, ATLAS9 model atmospheres \citep{Kurucz..1993,CastelliKurucz2003} and a linelist from the Vienna Atomic Line Database \citep[VALD;][]{Piskunov..VALD..1995,Ryabchikova..VALD..1997,Kupka..VALD..1999,Kupka..VALD..2000,Ryabchikova..VALD..2015}.

\begin{figure}
\includegraphics[clip=,trim=0.5cm 1.0cm 0.2cm 0.0cm,width=1.0\columnwidth]{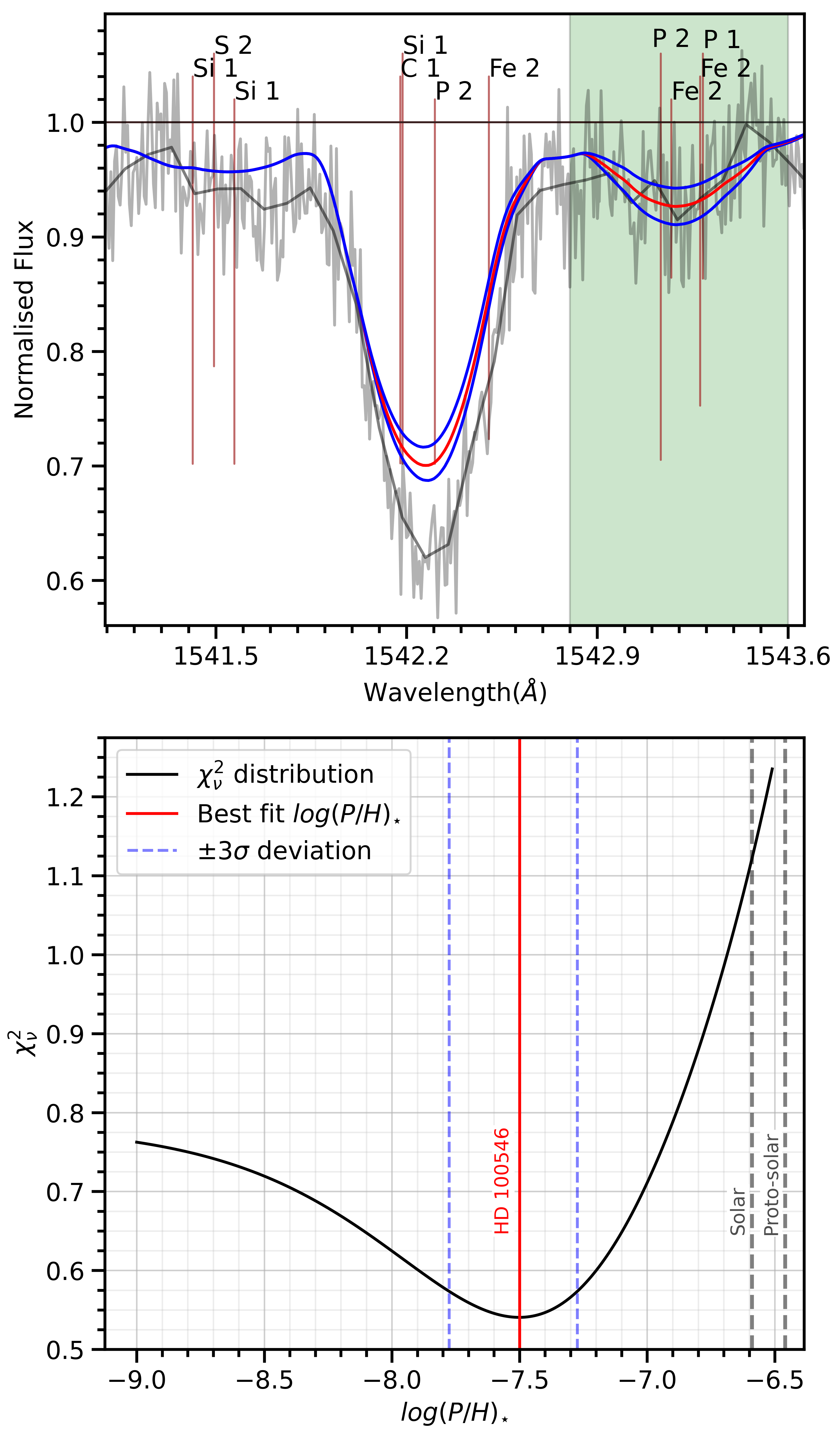}
\caption{\emph{\textbf{Top panel: }}The stellar spectrum of HD\,100546 is in light grey, the binned spectrum is in dark grey, the red spectrum is the best-fit synthetic spectrum, and the blue spectra are for $\pm3\,\sigma$ parameter deviations.
The highlighted region is used for estimating \stellarPH using the chi-square minimisation. \emph{\textbf{Bottom panel: }}$\chi^2$ as a function of \stellarPH. The red vertical line indicates the minimum, while the blue lines are $\pm3\,\sigma$ confidence intervals.}
\label{fig:stellarPH}
\end{figure}

The lines in the STIS UV spectrum of HD\,100546 are mostly blended due to crowding and rotational broadening ($v\sin{(i)}\approx65\,$\kms). Our analysis of the weak phosphorus lines between $1301$ and $1787\,$\AA\ is further complicated by the low SNR and by inaccuracies in UV line parameters. Based on a visual inspection of the spectrum, we chose a small wavelength region from $1542.8\,$\AA\ to $1543.5\,$\AA, dominated by two P lines, as shown in the top panel of Figure\,\ref{fig:stellarPH}. The stronger absorption feature at ${\approx}1542.5\,$\AA\ is a blend of a \ion{P}{ii} line with an equally strong carbon and slightly weaker silicon line. Our model of the strong \ion{P}{ii} feature likely suffers from errors in the \ion{C}{i} and \ion{Si}{i} line modelling. Adopting C/H and Si/H values from \citet{Kamaetal2016b}, we find the stronger \ion{P}{ii} absorption feature is too deep to be explained by an excess of phosphorus abundance. This is because the excess phosphorus leads to a significant over-prediction of the weaker \ion{P}{i}/\ion{P}{ii} line. 

The accuracy of oscillator strengths is a concern in particular when analysing UV spectra. The weaker absorption feature is a blend of a singly ionised \ion{P}{ii} line at ${\approx}1543.133\,$\AA\, a neutral \ion{P}{i} line at ${\approx}1543.288\,$\AA\, a \ion{Mn}{ii} line, and three \ion{Fe}{ii} lines. We tested the effect of the statistically weighted oscillator strength, \loggf, of the P and Fe lines, and found that \ion{P}{ii} dominates the weak feature. The \loggf\ values of the \ion{P}{ii} and \ion{P}{i} lines reported in the VALD database are -2.295 \citep{Kurucz..2012} and -0.820 \citep{Kurucz..Peytremann..1975}, respectively.

To determine the stellar photospheric phosphorus abundance, \stellarPH, for HD\,100546, we estimated the \loggf\ value for the \ion{P}{ii} line at ${\approx}1543.133\,$\AA\ by a $\chi^{2}$-minimisation fitting of the HST/STIS spectrum of Vega with stellar parameters and photospheric abundances from \citet{Fitzpatrick2010}. This author determined \stellarPH\ for Vega using an IUE spectrum with 78 lines of phosphorus. From our analysis of Vega, we obtained \loggf\,$=-2.224$ for the \ion{P}{ii} line at ${\approx}1543.133\,$\AA. We adopted this for our analysis of HD\,100546.


Using $\chi^{2}$ minimisation on a continuum-normalised spectrum, we find the stellar photospheric phosphorus abundance in HD\,100546 to be \stellarPH\,$=-7.50^{+0.23}_{-0.28}$ ($3\,\sigma$ confidence; see Figure\,\,\ref{fig:stellarPH}, right-hand panel). \resub{This result is validated with a fit on a second spectral region, which yields a \stellarPH\ value within $1\,\sigma$ of the above result (see Appendix\,\ref{apx:secondPH}).}

\subsubsection{Stellar parameters for the disk model}
We use the absolute stellar spectrum constructed by \citet{Brudereretal2012} using dereddened FUSE and IUE observations at UV wavelengths, and extended to longer wavelengths using the B9V template of \citet{Pickles_1998}. The stellar luminosity is $36\,L_\odot$ \citep{Kamaetal2016b}. The X-ray spectrum was characterized as a thermal spectrum with a temperature of $7\times10^7\,$K from $1$ to $100\,$keV, with a luminosity of $L_X = 7.94 \times 10^{28}\,$erg\,s$^{-1}$.

\subsection{Phosphorus in the protoplanetary disk gas }\label{sec:diskPH}

Next, we analyse the APEX data described Section\,\ref{sec:apexdata} to constrain the total gas-phase elemental phosphorus abundance in the HD\,100546 disk. We ran source-specific models using the 2D thermo-chemical code DALI \citep{Brudereretal2012, Bruderer2013}. Using an input stellar spectrum \citep{Brudereretal2012} and a 2D physical disk structure, the code iterates between continuum Monte Carlo radiative transfer and a chemistry solver to converge on the thermal and chemical structure of the disk, which allows to model outputs such as molecular lines. The disk structure we use for everything except the distribution of phosphorus species, is based on the HD\,100546 disk model consecutively improved by \citet{Brudereretal2012, Kamaetal2016b, Keyteetal2023, Keyteetal2024}. Below, we briefly restate some of the key features of the model. The adopted DALI model parameters are listed in Table\,\ref{tab:DALIparams}.

The disk structure is parameterised, with a surface density that follows the standard form of a power law with an exponential taper:
\begin{equation}
    \Sigma_\text{gas} = \Sigma_\text{c} \cdot \bigg(\frac{r}{R_c} \bigg)^{-\gamma} \exp \bigg[- \bigg(\frac{r}{R_c} \bigg)^{2-\gamma} \bigg]
\end{equation}
where $r$ is the radius, $\gamma$ is the surface density exponent, and the reference surface density at $R_\text{c}$ is $\Sigma_\text{c}/e$. The dust surface density is \sigmagas/\gasdust, where \gasdust is the gas-to-dust mass ratio. The vertical distribution of gas follows a Gaussian with scale height:

\begin{equation}
    h(r) = h_c\bigg(\frac{r}{R_c}\bigg)^\psi
\end{equation}
where $h_\text{c}$ is the scale height at $R_\text{c}$, and the power law index of the scale height, $\psi$, describes the flaring of the disk.

\sigmagas \ and \sigmadust \ extend from the dust sublimation radius ($R_\text{sub} = 0.25$ au) to the edge of the disk ($R_\text{out} = 500$ au), and can be varied independently inside the cavity radius \rcav \ with the multiplication factors \deltagas \ and \deltadust.

Dust settling is implemented by considering two different populations of grains: small grains (0.005-1 $\mu$m) and large grains (0.005-1 mm). The vertical density structure of the dust is such that large grains are settled towards the midplane, prescribed by the settling parameter, $\chi$\footnote{Note that the dust settling parameter $\chi$ is distinct from the $\chi^{2}$ statistic used elsewhere in the paper.}:

\begin{equation}
    \rho_\text{dust, large} = \frac{f \Sigma_\text{dust}}{\sqrt{2\pi}r \chi h} \exp \bigg[ -\frac{1}{2} \bigg( \frac{\pi/2 - \theta}{ \chi h} \bigg) ^{2} \bigg]
\end{equation}

\begin{equation}
    \rho_\text{dust, small} = \frac{(1-f)\Sigma_\text{dust}}{\sqrt{2\pi}rh} \exp \bigg[ -\frac{1}{2} \bigg( \frac{\pi/2 - \theta}{h} \bigg) ^{2} \bigg]
\end{equation}
where $f$ is the mass fraction of large grains and $\theta$ is the opening angle from the midplane as viewed from the central star. Both grain populations follow a size distribution prescribed by a power law with index $q=-3.5$.

\begin{figure}
 \includegraphics[width=\columnwidth]{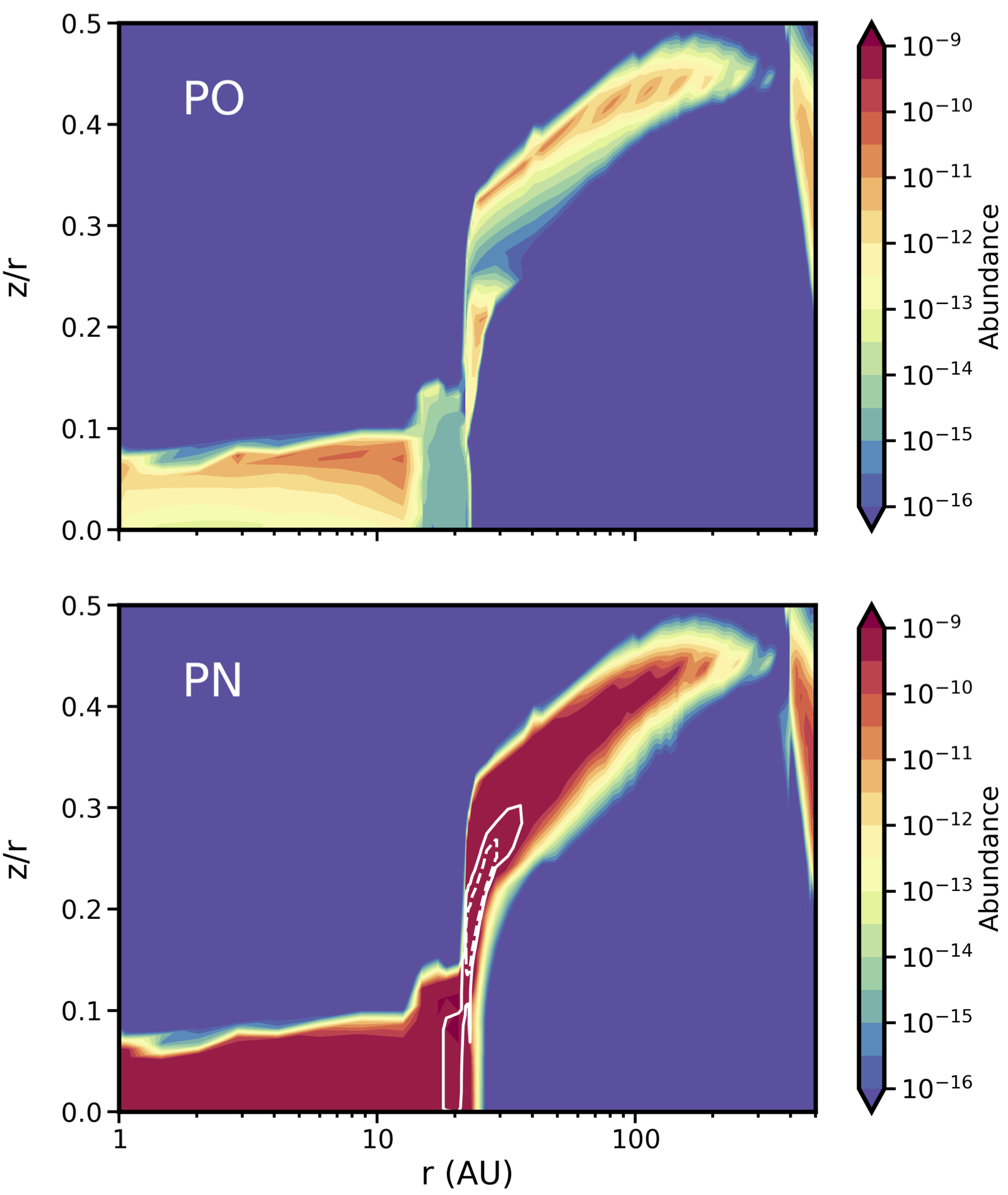}
 \caption{PO and PN abundance maps from our DALI model assuming a global volatile phosphorus abundance P/H $=10^{-9}$. The white contours enclose 50\,\% and 75\,\% of total line flux for the PN 234.936 GHz transition.}
 \label{fig:disk_Pabundances}
\end{figure}

\subsubsection{Chemical network}
The chemical reaction network used in our model is based on a subset of UMIST\,06 \citep{woodall2007}. It consists of 145 species (including neutral and charged PAHs) and 1814 individual reactions. The DALI code includes H$_2$ formation on dust, freeze-out, thermal desorption, hydrogenation, gas-phase reactions, photodissociation and -ionization, X-ray induced processes, cosmic-ray induced reactions, PAH/small grain charge exchange/hydrogenation, and reactions with vibrationally excited H$_2$. Non-thermal desorption is only included for a small number of species (CO, CO$_2$, H$_2$O, CH$_4$, NH$_3$). For grain surface chemistry, only hydrogenation of simple species is considered (C, CH, CH$_2$, CH$_3$, N, NH, NH$_2$, O, OH). The details of these processes are described in \citet{Brudereretal2012}. For the present study, we have updated the network with 26 phosphorus-bearing species and 135 reactions from UMIST. Binding energies for all P-bearing species are set to $E_{\rm B}=5770$ K following the values determined for PO and PN by \citet{piacentino_2022}, equivalent to a sublimation temperature of $T \approx 125\,$K. The elemental abundances used in our model are listed in Table \ref{tab:DALIabuns}.

\begin{figure}
 \includegraphics[width=\columnwidth]{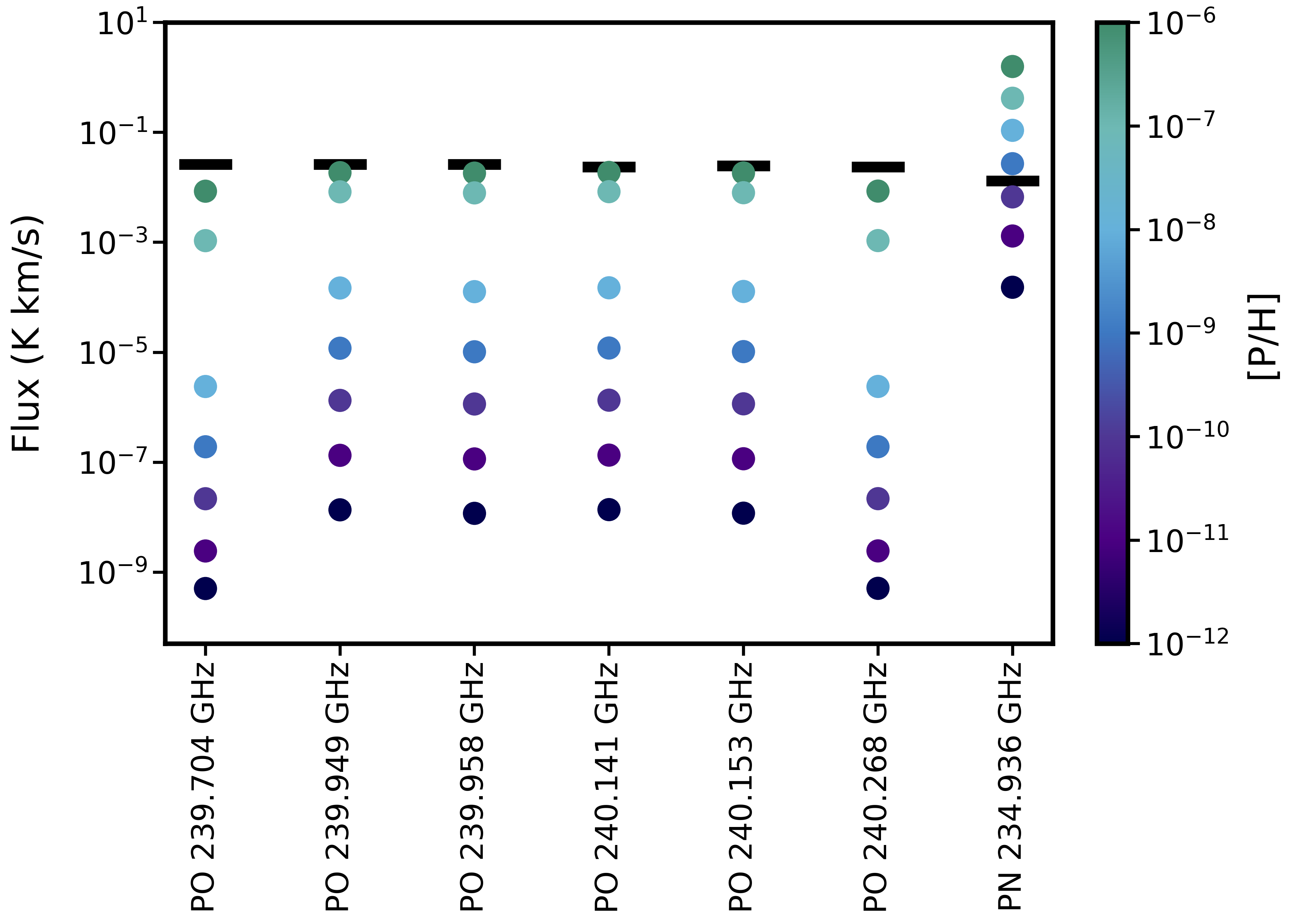}
 \caption{Volatile elemental P/H in the HD\,100546 disk assuming a globally constant abundance. Disk-integrated $3\sigma$ line flux upper limits for the PO and PN transitions observed with APEX (black lines), and predicted line fluxes for a DALI model grid covering volatile elemental phosphorus abundances from P/H\,$=10^{-12}$ to $10^{-6}$ (colored circles) are shown.}
 \label{fig:disk_Pfluxes}
\end{figure}

\begin{figure*}
 \includegraphics[width=0.80\linewidth]{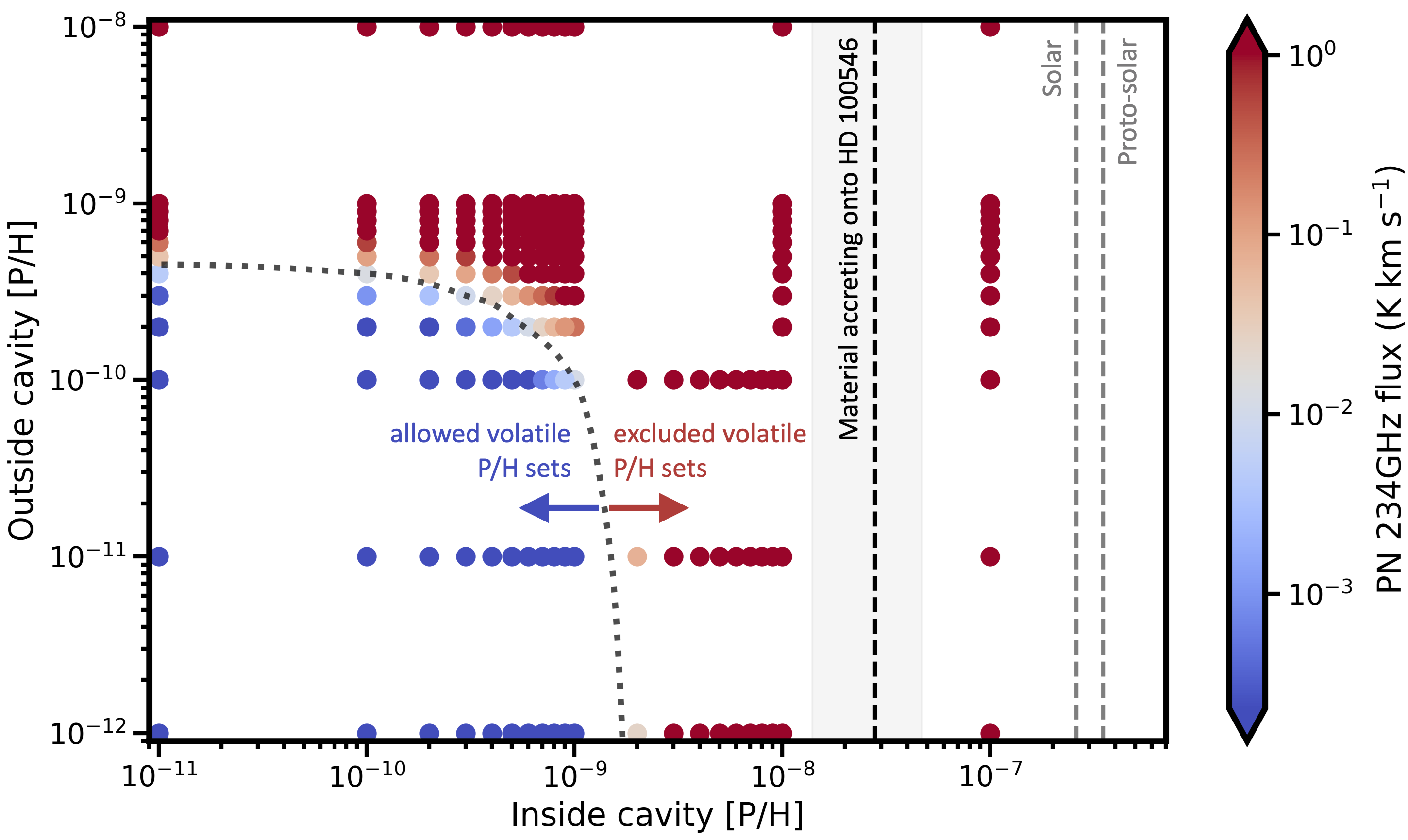}
 \caption{Coupled constraints on the gas-phase elemental P/H ratio inside (P/H$_{\rm in}$, $x$-axis) and outside  (P/H$_{\rm out}$, $y$-axis) the $22\,$au dust cavity in the HD\,100546 disk. Allowed abundance sets (shades of blue) meet the constraint from the PN line disk-integrated $3\sigma$ flux upper limit. The inner cavity P/H value is compared with the ratio as measured on the star for the accreting material, (P/H)$_{\star}$, and the solar \citep{Asplundetal2009} and proto-solar \citep{Lodders2003} values.}
 \label{fig:diskstar_inout}
\end{figure*}

\subsubsection{Molecular excitation and transition data}

We adopted data for PO and PN rotational energy levels and transitions from the LAMDA database \citep{Schoieretal2005, tobola_2007, lique_2018}. Some necessary corrections to the energy level ordering were made by J.Bergner.

\subsubsection{Gas-phase elemental P/H in the disk}

To constrain the \emph{volatile }elemental phosphorus abundance in the disk, we used a two-stage modelling approach:

Firstly, \resub{to get a rough estimate of the global depletion of gas-phase phosphorus,} we modelled the disk-integrated PO and PN line flux upper limits using a constant volatile P/H ratio across the whole disk. Representative gas-phase PO and PN number abundance maps can be seen in Figure\,\ref{fig:disk_Pabundances}. The modelled line fluxes, \resub{with volatile} elemental phosphorus values from P/H\,$=10^{-12}$ to $10^{-6}$ (solar-like), are  shown in Figure\,\ref{fig:disk_Pfluxes}. The PO lines provided a weak constraint on the volatile phosphorus: P/H\,$\lesssim 10^{-6}$, which is an order of magnitude higher than the solar value. Our globally constant total elemental gas-phase P/H is based on the PN\,$6-5$ line, from which we find (P/H)$_{\rm global}\lesssim 5\times 10^{-10}$.

Secondly -- and this is our final, best-fit model -- we modelled the data by dividing the disk into two radial zones with an independent gas-phase P/H ratio: one for material inside the inner edge of the main dust ring (i.e., inside the $22\,$au dust cavity; P/H$_{\rm in}$), and another for the outer disk (P/H$_{\rm out}$). \resub{This two-zone model was motivated by the abrupt change in temperature near the edge of the large dust cavity, with the temperature reaching $\approx180\,$K at the inner edge of the dust ring \citep{Keyteetal2023}; and by the assumption that inward-drifting pebbles would stop in the dust ring and release their volatile ices. }The results are shown in Figure\,\ref{fig:diskstar_inout} and show that a useful set of upper limits, described by a surface in the (P/H)$_{\rm in}$, (P/H)$_{\rm out}$ coordinate space, can be obtained. Looking at the extreme values for both axes, we independently constrain the gas-phase total P/H ratio inside the dust ring to be P/H$_{\rm in}\lesssim2\times10^{-9}$, while outside the dust ring we find P/H$_{\rm out}\lesssim5\times10^{-10}$.

\resub{Since the two protoplanet candidates are located inside (planet c) and outside (planet b) the main dust ring, the P/H$_{\rm in}$ and P/H$_{\rm out}$ limits can, respectively, be associated with gas currently being accreted by the protoplanets.}

\section{Discussion}\label{sec:discussion}

We have identified that, in the HD\,100546 system, the total elemental gas-phase phosphorus abundance is depleted inside the main dust ring by a factor $\geq129$ and outside this zone by a factor $\geq513$ (obtained assuming a solar reference abundance and dividing that by the inner and outer disk upper limits from Section\,\ref{sec:diskPH}). We also find that the total (gas \emph{plus} solid) elemental P is depleted in the dust-depleted disk material accreting onto the star by a factor $\approx8$ (obtained from dividing the solar reference by the \stellarPH\ from Section\,\ref{sec:stellarPH}).  These results provide independent insights into the behaviour of P in the HD\,100546 system, which we discuss below. 

\begin{figure*}
\includegraphics[clip=,width=0.8\linewidth]{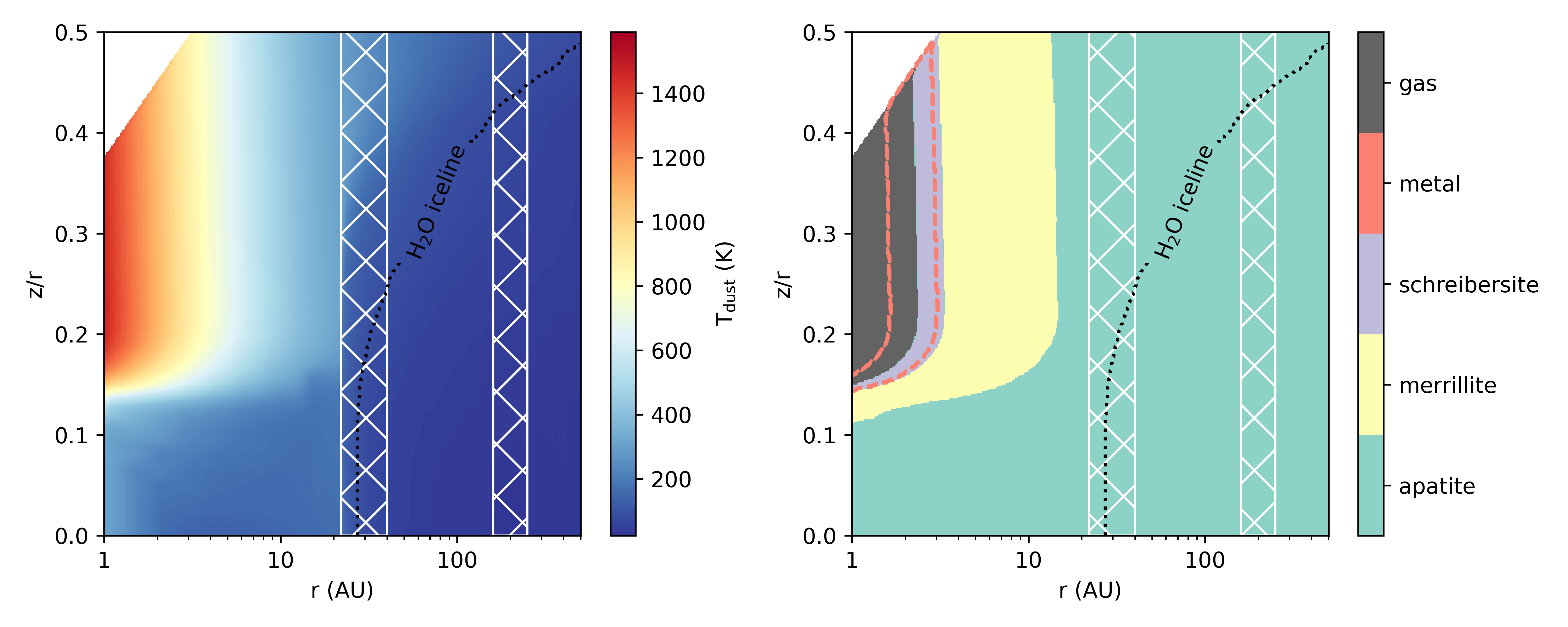}
\caption{The dominant phosphorous reservoirs throughout the HD\,100546 disk.  \emph{Left-hand panel: }a map of the calculated disk temperature structure, which dictates phosphorous mineral stability.  Normalised height above midplane is given by z/r, with height above midplane, z\,(au), divided by the radial distance, r\,(au).  Hatched regions indicate the two dust rings identified in this system and the dotted black line marks the water iceline (parameterised simply as the 121\,K isotherm after \citealt{Lodders2003}).  \emph{Right-hand panel: }a map of the dominant phosphorous reservoirs in the disk assuming equilibrium at the dust temperature.  The minerals apatite, merrillite, and schreibersite progressively replace each other as the main phoshporous reservoir as temperature is increased, up to the point where temperatures are high enough that all major phosphorous-bearing mineral phases have sublimated and most phosphorous is in the gas phase.  Phosphorous dissolved in Fe metal is never a dominant  reservoir, but its presence is mapped by the dashed contour.  Mineral stabilties were taken from \citet{pasek2019_icarus}.}
\label{fig:min_map}
\end{figure*}

\begin{figure*}
\includegraphics[clip=,width=1.0\linewidth]{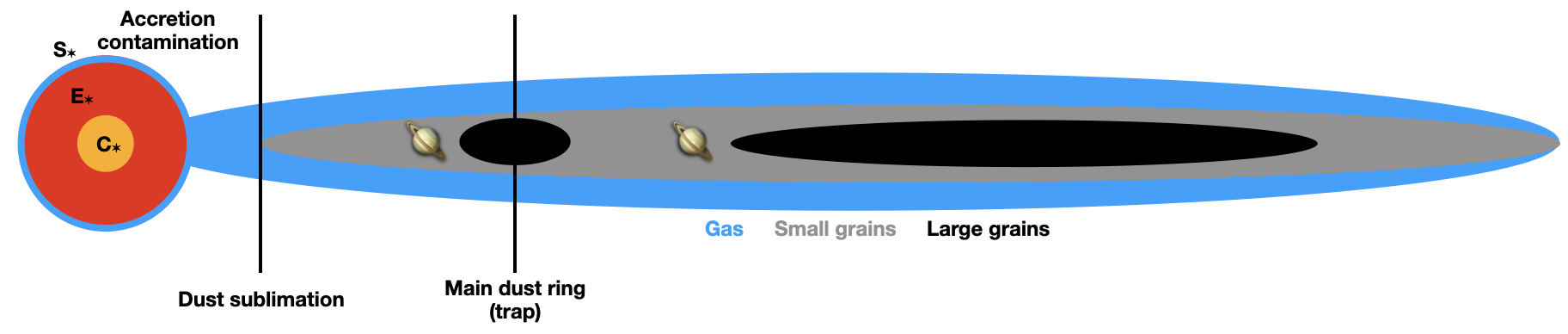}
\caption{Sketch of the HD\,100546 system with the star, protoplanetary disk, and two claimed protoplanets.}
\label{fig:hd100546_cartoon}
\end{figure*}

\subsection{Phosphorus volatility in the HD\,100546 disk}

The factor of $8$ depletion of phosphorus on the stellar surface, identified above (see Section\,\ref{sec:stellarPH}), is consistent with the factor of $\sim10$ overall depletion of refractory elements on the stellar surface \citep{Kamaetal2016b}.

\resub{In addition to a factor of $\approx8$ depletion of P, the accretion-dominated photosphere of the star HD\,100546 is depleted in other refractory elements, also by approximately a factor of $10$ \citep[e.g., Mg, Ca, Si, Al; see Appendix\,A in][]{Kamaetal2016b}.  This is despite the disk itself having an overall a gas-to-dust ratio of $\Delta_{\rm /gd}=10$--$100$ \citep[e.g.,][]{Brudereretal2012, Kamaetal2016b}.  Together, these observations imply that approximately $90\,$\% of the inward-moving dust mass from the outer disk does not currently reach the star.}


The most significant finding from the two-zone disk model is that the total gas-phase phosphorus abundance in the inner cavity is P/H$_{\rm in}<2\times10^{-9}$, which is at least $16$ times less than the abundance on the stellar photosphere. We identify the stellar P/H$_{\star}$ value with the total gas \emph{plus} solid abundance of phosphorus in the inner cavity. In other words, P/H$_{\star}$ represents any and all carriers of phosphorus that are carried past the main dust ring and through the inner cavity. Given the large difference between P/H$\star$ and P/H$_{\rm in}$, the two-zone disk model strongly suggests that most of the phosphorus passing through the dust cavity is contained in small dust grains, hidden from the gas-phase chemistry probed by our ALMA observations.

The values of P/H$_{\rm star}$ and P/H$_{\rm in}$ are thus consistent with a scenario where most of the inward-moving dust mass has gotten trapped, most likely in the prominent dust ring at $22$ to $40\,$au, leaving the inner cavity deprived of large dust grains (pebbles) and the stellar surface deprived of dust-forming elements. Only small dust grains, which are coupled to the gas but generally constitute a small fraction of the total dust mass, can traverse the inner cavity. \resub{In the HD\,100546 disk model from \citet{Keyteetal2024}, grains $\leq1\,$mm in size constitute $5\,$\% of the dust mass, which given the various uncertainties involved appears consistent with $\approx10\,$\% of refractories reaching the stellar surface.} This general finding on refractories being depleted on the star due to dust trapping is in line with our previous work in \citet{Kamaetal2016b, Kamaetal2019}. The stellar and inner disk P/H values furthermore lead to the conclusion that the dominant phosphorus reservoir in the HD\,100546 disk is \resub{sufficiently refractory to remain solid at the inner edge of the main dust belt, where the temperature reaches nearly $200\,$K.}

We are here using ``refractory'' more broadly compared to its cosmochemical meaning, and use it to refer to anything less volatile than water ice.  This distinction is made on the basis that the stellar oxygen abundance is normal, i.e. undepleted, in HD\,100546 and other Herbig\,Ae/Be stars with dust-poor accretion.  This constant O/H value regardless of whether the disk is transitional or full, is taken as evidence that H$_{2}$O ice, which is the dominant carrier of elemental O in disks, generally sublimates in the main dust trap in such transitional disks and freely accretes onto the star \citep{Kamaetal2019}. Ice sublimation is in general more likely at the inner edge of dust traps, as temperatures rise due to increased illumination  \citep{broome2023_mnras, Chenetal2024}. In the case of HD\,100546, the lack of oxygen depletion is consistent with the matching locations of the water iceline and inner dust ring \citep[][see also Figure\,\ref{fig:min_map}]{Keyteetal2023, Keyteetal2024}. 

\resub{We note that while the volatility trend of elemental depletions on the surface of accreting Herbig\,Ae/Be stars is consistent from the most volatile elements (C), through the intermediates (Na, S, Zn), to the most refractory ones (e.g., Fe, Ti), oxygen does not follow this trend \citep{Kamaetal2015b, Kamaetal2019}. Instead, in material accreting onto Herbig\,Ae/Be stars oxygen has a  refractory fraction  of $0\pm2\,$\% \citep{Kamaetal2015b}. This may be due to with disk or stellar evolution processes, or systematics in the oxygen abundance measurements and is being investigated in a separate study. While this raises doubts as to the overall reliability of the oxygen abundance, we argue that the effect is likely mainly relevant for masking the minor refractory component of O. If H$_{2}$O ice, the expected major O carrier, was being trapped in the last major dust trap, the star would be receiving strongly oxygen-depleted material, which would then show a depletion signature of comparable magnitude to Fe or Ti.} 



This finding of elemental P being locked almost exclusively in dust is consistent with the temperature at the location of the dust trap, which predicts that -- at equilibrium, which is a limiting assumption -- most phosphorous at this location would be in refractory mineral form as apatite (Figure\,\ref{fig:min_map}). \resub{We only include the equilibrium composition as a secondary argument, to suggest that our findings are not inconsistent with the P reservoirs expected in case the composition had time to equilibrate in the dust trap.} Apatite or large phosphorus oxides (e.g.~P$_4$O$_{10}$) with similar volatility were also inferred to be likely grain-phase carriers of phosphorus in the protostellar stage \citep{Bergner2022}.



\subsection{The main phosphorus reservoir}

The conclusion from above that most, or all, elemental P is locked in refractory solids. This solid P is most likely trapped in pebbles in the $22$--$40\,$au dust ring, similar to even more refractory elements such as iron, titanium, or aluminium. The accreted oxygen abundance on the HD\,100546 stellar photosphere is consistent with almost no depletion, indicating almost all oxygen passes freely through the dust trap. If we assume the elemental O budget is dominated by H$_{2}$O, we can conclude the main reservoir of elemental phosphorus in the HD\,100546 disk is more refractory than H$_{2}$O ice, so that it can remain trapped while H$_{2}$O ice sublimates. \resub{The temperature of the dust trap is $\approx70\,$K (outer edge) to $\approx180\,$K (inner edge), sufficient to sublimate water \citep{Kamaetal2016b, Keyteetal2023}.} 


Our empirical finding that almost all phosphorus is locked in solids that are more refractory than water ice is consistent with predictions from thermochemical equilibrium. As seen in Figure\,\ref{fig:min_map}, we find that the preferred equilibrium reservoir is a mineral phase in much of the disk, and specifically in the inner dust ring. Equilibrium condensation sequence calculations place P in the refractory minerals schreibersite, apatite, or merrillite \citep{FegleyLewis1980, Lodders2003, Pirimetal2014, pasek2019_icarus}. We note that it is unclear whether thermochemical equilibrium is an appropriate assumption for the material in the HD\,100546 dust ring.

Among meteorites, the main P reservoir in hydrous carbonaceous chondrites and ordinary chondrites is phosphates, such as apatite ($\geq90\,$\% of total elemental phosphorus), or schreibersite \citep[$\leq10\,$\%;][and references therein]{PasekLauretta2008}. However, the \emph{Rosetta} mission to comet 67P/Churyumov-Gerasimenko found no evidence for apatite and only weak consistency with schreibersite for the main P carrier in the cometary dust grains \citep{Gardneretal2020}. \resub{In samples from asteroid Ryugu, the main phosphorus-bearing phases were found to be apatite, schreibersite, and a new phase: hydrated ammonium-magnesium phosphorus (HAMP) grains \citep{Pilorgetetal2024}. A similar list of P-carriers has been found in samples from asteroid Bennu \citep{Laurettaetal2024}.}


Ammonium \resub{salts, such as the ammonium phosphate found recently on Ryugu (see above),} have become a contender for carrying a significant fraction of nitrogen, sulfur, or other elements such as halogens in a semi-refractory reservoir, sublimating at a few hundred kelvins, based on solar system evidence \citep{Pochetal2020, Altweggetal2020, Altweggetal2022}. While we cannot rule out ammonium phosphate, (NH$_{4}$)$_{3}$PO$_{4}$, as the main carrier of phosphorus in \resub{the HD\,100546 disk, it we also have no evidence for a significant total mass of diverse ammonium salts being withheld} in the dust trap. This is because the accreted photospheric N/H ratio of HD\,100546, $\log{\rm(N/H)_{\star}}=-3.82\pm0.30$ \citep{Kamaetal2016b}, is marginally super-solar, assuming$\log{\rm(N/H)_{\odot}}=-4.17\pm0.05$ \citep{Asplundetal2009}. Thus, we have no evidence for nitrogen depletion in the accretion stream. 

\resub{On the other hand,} the small fraction of elemental N needed to lock away only P atoms in ammonium phosphate would be too small to cause a measurable N depletion on the star ($3\times$P/N\,$\approx 1\,$\% for solar abundances) so we cannot rule it out. However, the lack of evidence of trapping of more abundant salt species in the dust trap then demands a very specific explanation. If ammonium phosphate were an important P reservoir, a mechanism strongly favouring its formation over NH$_{4}^{+}$ salts with a more abundant anion (such as sulfate) would be required to explain the lack of measurable elemental N depletion in the HD\,100546 stellar photosphere. 

\resub{Our available evidence therefore favours a mineral reservoir, such as apatite or schreibersite, to account for most of the phosphorus. Ammonium phosphate is also a candidate, but it may not have a dominant role as we do not see a depletion of N that would indicate the presence of large amounts of diverse ammonium salts as a ``volatile vault'' in the dust belt where P is getting trapped.}

\subsection{The phosphorus budget of the protoplanets}

As described in Section\,\ref{sec:target}, the disk around HD\,100546 is known to host at least two giant protoplanet candidates: HD\,100546\,b \citep[$a_{\rm maj,b}=143\,$au, $M_{\rm b}=3\,$M$_{\rm Jup}$][]{} and c \citep[$a_{\rm maj,b}=13\,$au,$M_{\rm b}=8\, $M$_{\rm Jup}$][]{Pinillaetal2015, Pyerinetal2021}. We illustrate the scenario for HD\,100546 in Figure\,\ref{fig:hd100546_cartoon}. Protoplanet b orbits in a wide dust gap radially outwards from the main dust trap (ring), while c orbits inwards of the trap \citep{Walshetal2014, Pinillaetal2015}. 

Our measurements above show that elemental P in the inner few tens of au of this disk can be accounted for entirely by a refractory reservoir: the depletion of phosphorus from the gas follows that of Fe, Mg, Si, and other refractories \citep[see also][]{Kamaetal2015b}. In an imaginary parcel of homogeneously mixed disk material, almost all P would be contained in dust grains. Most of the dust mass is in large dust grains (pebbles) and thus settled on the midplane, where such large grains will drift into dust traps such as the main dust ring. A smaller fraction of total \resub{elemental }P will be in small grains, which are kinematically coupled to the gas and not as sensitive to trapping in local gas pressure maxima.

The above implies the current availability of phosphorus may be very different for the protoplanets b and c. A giant planet may accrete material into its gas envelope either across the disk midplane, or through the surface via meridional flows \citep{Teagueetal2019, Szulagyietal2022}. As both planets have opened dust gaps, any currently ongoing midplane accretion would be dust-poor as large grains would not be able to accrete. Furthermore, we can hypothesise the inner planet (c) might be accreting particularly low-metallicity material as any solids would need to have first passed through \emph{two }radial dust traps, with the inner trap apparently being the stronger of the two. If this is the case, planet c may end up with a lower metallicity, and a lower total elemental P/H ratio in its envelope compared to the outer planet, b. 

If meridional flows dominate the currently ongoing accretion onto either of the planets, that planet (or both) would have a low envelope metallicity, with a correspondingly low elemental P/H ratio.

\section{Conclusions}

\begin{enumerate}
\item{We find a phosphorus abundance $\log{(P/H)_{\star}=-7.50^{+0.23}_{-0.28}}$ in the stellar photosphere. This is significantly below the solar value, $\log{(P/H)_{\odot}=-6.59\pm0.03}$. The slowly-mixing radiative envelope of the star and the high disk accretion rate imply the (P/H)$_{\star}$ value \resub{characterises the total P/H in the inner disk, where the accretion originates, and not the bulk of the star}.}
\item{We find the total gas-phase elemental phosphorus abundance in the inner disk, inside the main dust ring, to be (P/H)$_{\rm in}\lesssim 2\times10^{-9}$. We constrain the value outside the dust ring to be (P/H)$_{\rm out}\lesssim 5\times10^{-10}$. \resub{These two values characterise P/H in the gas currently being accreted by protoplanet candidates c and b, respectively.}}
\item{Phosphorus in the HD\,100546 disk is strongly depleted from the gas into a refractory (dusty) reservoir. This is based on the above conclusions that phosphorus is strongly depleted from the disk gas everywhere, and depleted -- but less severely -- in the material (gas$+$dust) in the accretion stream. The main carrier of P in the inner cavity is predicted to be small dust grains that are not trapped in the dust ring at $22$ to $40\,$au.}
\item{Apatite and schreibersite are favoured over volatile ices or ammonium salts as the dominant phosphorus reservoir in the grains, both on sublimation temperature grounds and on the grounds that we see the volatiles N, C, and O behaving differently from P in the inner disk.}
\item{Our empirical finding that elemental phosphorus in the HD\,100546 disk is carried entirely, or almost entirely, by dust suggests that the P/H ratio in giant planet envelopes traces their refractory element accretion history as well or even better than sulfur.}
\end{enumerate}

\section*{Acknowledgements}
The authors thank Anish Amarsi for useful comments that helped to improve the manuscript. SB, CF, LF, and MK gratefully acknowledge funding from the European Union's Horizon Europe research and innovation programme under grant agreement No. 101079231 (EXOHOST), and from UK Research and Innovation (UKRI) under the UK government’s Horizon Europe funding guarantee (grant number 10051045). LK acknowledges funding via a Science and Technology Facilities Council (STFC) studentship and by UKRI guaranteed funding for a Horizon Europe ERC consolidator grant (EP/Y024710/1).

\section*{Data Availability}

The sub-millimetre spectroscopic observations of rotational lines of PO, PN, and HCP can be accessed on the APEX data archive\footnote{https://www.apex-telescope.org/ns/}, under the proposal code 0102.C-0927(A). The \emph{Hubble }Space Telescope spectra are available through the StarCAT database\footnote{\texttt{https://archive.stsci.edu/prepds/starcat/StarCAT\_portal.html}}. \resub{Our disk model is available on request from the authors.}

\bibliographystyle{mnras}
\bibliography{puplim.bib} 


\appendix
\section{Disk model parameters}
In Table\,\ref{tab:DALIparams}, we list the DALI model parameters adopted for the HD\,100546 disk model. The elemental abundances assumed for elements other than P are listed in Table\,\ref{tab:DALIabuns}.

\begin{table}
\caption{DALI disk model parameters for HD\,100546 \citep{Keyteetal2023}. See also \citet{Brudereretal2012} and \citet{Kamaetal2016b}.}
\label{tab:DALIparams}      
\centering
\begin{tabular}{l l l}     %
\hline\hline       
                      
Parameter & Description & Fiducial\\ 
\hline                    
   \rsub                & Sublimation radius                         & 0.25 au    \\
   \rgap                & Inner disk size                            & 1.0 au     \\
   \rcav                & Cavity radius                              & 22 au      \\
   $R_\text{out}$       & Disk outer radius                          & 1000 au    \\
   $R_c$                & Critical radius for surface density        & 50 au      \\
   \deltagas            & Gas depletion factor inside cavity         & $10^{-5}$  \\
   \deltadust           & Dust depletion factor inside cavity        & $10^{-4}$  \\
   $\gamma$             & Surface density power law index            & 1.0        \\
   $\chi$               & Dust settling parameter                    & 0.2        \\
   $f$                  & Large-to-small dust mixing                 & 0.95       \\
   $\Sigma_c$           & $\Sigma_\text{gas}$ at $R_c$               & 87.25 g cm$^{-2}$  \\
   $h_c$                & Scale height at $R_c$                      & 0.10       \\
   $\psi$               & Power law index of scale height            & 0.25       \\
   \gasdust             & Gas-to-dust mass ratio                     & 100        \\
   $L_*$                & Stellar luminosity                         & $36\; L_\odot$        \\
   $L_X$                & Stellar X-ray luminosity                   & $7.94 \times 10^{28} \text{ erg s}^{-1}$ \\
   $T_X$                & X-ray plasma temperature                   & $7.0 \times 10^{7}$ K     \\
   $\zeta_\text{cr}$    & Cosmic ray ionization rate                 & $3.0 \times 10^{-17}$ s$^{-1}$  \\
   $M_\text{gas}$       & Disk gas mass                              & $9.89 \times 10^{-2}$ \msun   \\
   $M_\text{dust}$      & Disk dust mass                             & $1.06 \times 10^{-3}$ \msun   \\
   $t_\text{chem}$      & Chemical timescale                         & \text{5 Myr} \\
\hline                  
\end{tabular}
\end{table}

\begin{table}
\centering
\caption{Chemical element number abundances, relative to H nuclei, in the DALI model of the disk around HD\,100546.}
\begin{tabular}{l l}     %
\hline\hline                     
Species & $n_i/n_{\rm H}$ \\ 
\hline
\multicolumn{2}{c}{\emph{Fixed}}\\
\hline
   H  & 1.0  \\
   He & $9.0 \times 10^{-2} $ \\
   C  & $1.5 \times 10^{-5} $ \\
   N  & $6.2 \times 10^{-5} $ \\
   O  & $3.0 \times 10^{-5} $ \\
   S  & $1.0 \times 10^{-8} $ \\
   Mg & $5.0 \times 10^{-7} $ \\
   Si & $5.0 \times 10^{-7} $ \\
   Fe & $5.0 \times 10^{-7} $ \\
\hline
\end{tabular}
\label{tab:DALIabuns}
\end{table}
\section{(P/H)$_{\star}$ from a second spectral region}
\label{apx:secondPH}

\begin{figure}
\includegraphics[clip=,width=1.0\columnwidth]{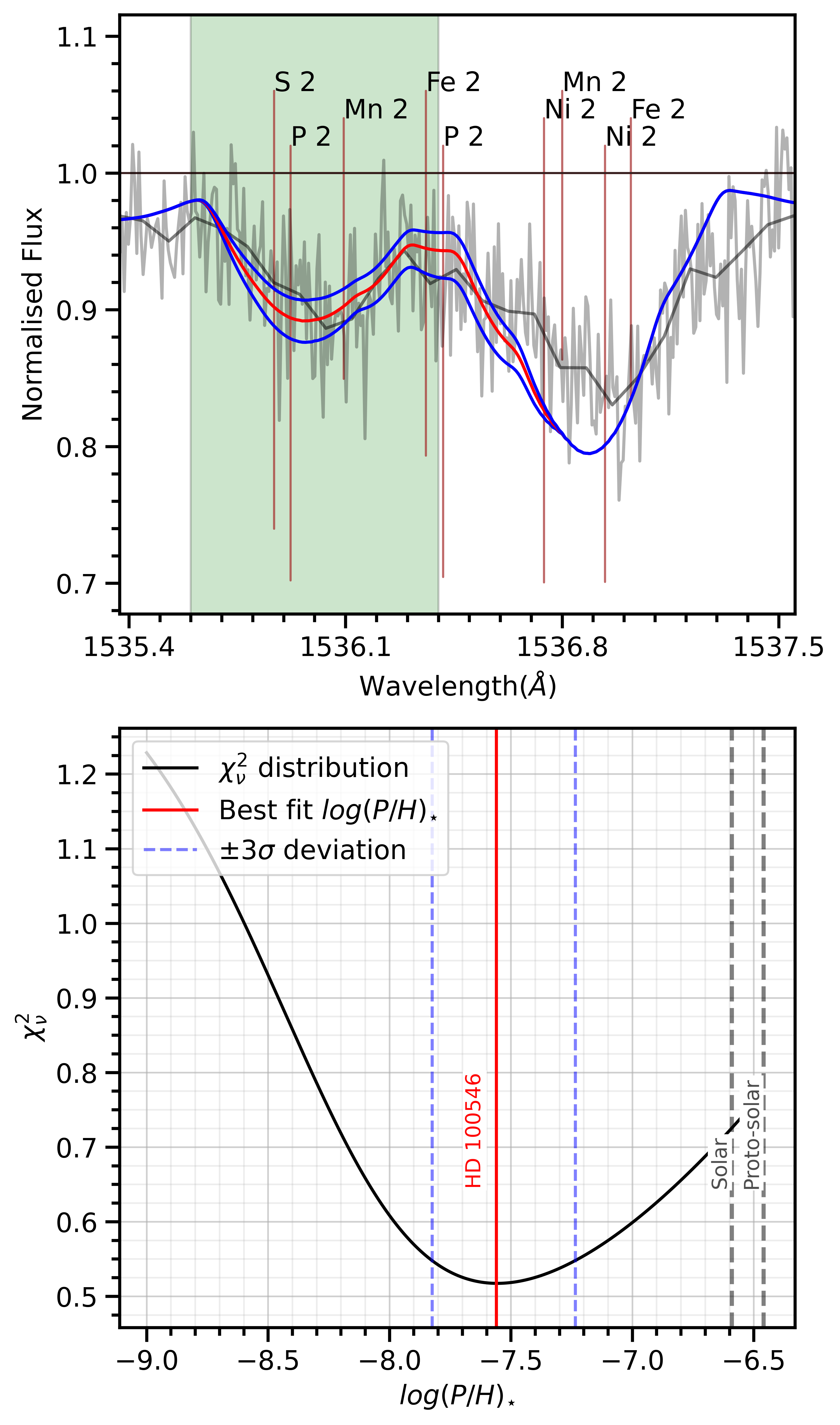}
\label{fig:UV second spectral region}
\caption{Determining (P/H)$_{\star}$ for HD\,100546 in a second spectral region. \emph{\textbf{Top panel: }}The spectrum of HD\,100546 at native (light grey) and binned resolution (dark grey). Overlays show the best-fit synthetic spectrum (red) and its range for $\pm3\,\sigma$ parameter deviations (blue). The highlighted green region is used for estimating \stellarPH\ with chi-square minimisation. \emph{\textbf{Bottom panel: }}$\chi^2$ as a function of \stellarPH. The red vertical line indicates the minimum (best fit), while the blue dashed lines are $\pm3\,\sigma$ confidence intervals.}
\end{figure}

\resub{
We also estimated the P abundance from the \ion{P}{ii} line between $1535.4$ and $1536.4\,$\AA. The \ion{P}{ii} line is blended with one \ion{S}{ii} and one \ion{Mn}{ii} line. No prior abundance measurement for Mn was available from \citet{Folsometal2012}, so we measured it from the following spectral regions: $1857.52$--$1859.13\,$\AA, $1863.57$--$1865.29\,$\AA, $1865.29$--$1870.33\,$\AA, $1910.00$--$1912.00\,$\AA, $1942.90$--$1945.95\,$\AA, and $1916.70$--$1922.00\,$\AA. We find (Mn/H)$_{\star}=-7.56 \pm 0.35$. We applied this value to estimate (P/H)$_{\star}$ from the spectral region highlighted in Figure\,\ref{fig:UV second spectral region}. By comparing with the Vega spectrum, we estimated the statistically weighted oscillator strength value for the \ion{P}{ii} line at $1535.9225\,$\AA\ to be \loggf$=-1.470$. The value for the \ion{P}{ii} line reported in VALD is \loggf$=-1.765$. There is another \ion{P}{ii} line at $1536.4157\,$\AA, which is not considered while fitting the spectra because the line is blended with two relatively strong \ion{Ni}{ii} lines which likely have incorrect \loggf\ values, based on a comparison with the Vega spectrum. We could not estimate the \loggf\ values for the other \ion{P}{ii} line and the two \ion{Ni}{ii} lines. Using the $\chi^{2}$ minimisation technique described in Section\,\ref{sec:stellarPH}, we estimated the P abundance from this spectral region to be (P/H)$_{\star}=-7.56^{+0.32}_{-0.26}$ ($3\,\sigma$ confidence; see Figure\,\,\ref{fig:UV second spectral region}, bottom panel). The best-fit value here is within $1\,\sigma$ of the main result presented in Section\,\ref{sec:stellarPH}.
}

\resub{
Aside from noise in the spectrum, two other sources of uncertainty affect our P abundance measurements - uncertainty in our \loggf\ values and the uncertainty in our continuum placement. For weak lines, the \loggf\ of a spectral line is directly proportional to the inferred abundance of that element. Since our correction of \loggf\ values for the P lines are based on the spectrum of Vega, we can use the P abundance uncertainty in Vega from \citet{Fitzpatrick2010} to get the uncertainty in our \loggf. This gives an uncertainty of 0.05\,dex in \loggf. The \loggf\ value can also be affected by the S/N of the Vega spectrum. This uncertainty is $\sim0.06$\,dex. Assuming both uncertainties are uncorrelated, the combined uncertainty of the \loggf\ value is $\sim0.08$\,dex. Thus, the uncertainty in our P abundance estimates of HD\,100546 from \loggf\ is 0.08\,dex ($1\,\sigma$). 
}

\resub{
For the spectral regions discussed here and in Section\,\ref{sec:stellarPH}, we shifted the continuum vertically to check its effect on the measured (P/H)$_{\star}$. We found that a local continuum multiplier in excess of a factor of 1.005 makes the local spectral region inconsistent with the wider spectrum in terms of the low-order behaviour of the normalised spectrum. A continuum multiplier $1.005$ in either direction for both spectral regions used to estimate (P/H)$_{\star}$ changes the phosphorus abundance by $\approx \pm$ 0.11 dex. This is smaller than the formal $3\sigma$ uncertainties we report for our results.
}


\resub{
We did not consider here the uncertainty due to other stellar parameters and abundances since quantifying that is more complicated and was not considered as a source of uncertainty in \citet{Folsometal2012}. Although there might be additional sources of systematic uncertainties, such as incorrect \loggf\ values of neighbouring blended lines, our comparison with the Vega spectrum suggests that such factors cannot offset the P depletion implied by our measurement.
} 

\bsp	
\label{lastpage}
\end{document}